\documentstyle[preprint,eqsecnum,aps]{revtex}
\newif\iftightenlines\tightenlinesfalse
\tightenlines\tightenlinestrue
\begin{document}
%
\def\pT{p_T^{\phantom{7}}}
\def\MW{M_W^{\phantom{7}}}
\def\ET{E_T^{\phantom{7}}}
\def\bh{\bar h}
\def\lm{\,{\rm lm}}
\def\lo{\lambda_1}
\def\lt{\lambda_2}
\def\pslt{p\llap/_T}
\def\eslt{E\llap/_T}
\def\to{\rightarrow}
\def\Re{{\cal R \mskip-4mu \lower.1ex \hbox{\it e}}\,}
\def\Im{{\cal I \mskip-5mu \lower.1ex \hbox{\it m}}\,}
\def\SU{SU(2)$\times$U(1)$_Y$}
\def\te{\tilde e}
\def\tt{\tilde t}
\def\tb{\tilde b}
\def\ttau{\tilde \tau}
\def\tg{\tilde g}
\def\tga{\tilde \gamma}
\def\tnu{\tilde\nu}
\def\tl{\tilde\ell}
\def\tq{\tilde q}
\def\tw{\widetilde W}
\def\tz{\widetilde Z}
\def\tx{\widetilde \chi}
%
%
\preprint{\vbox{\baselineskip=14pt%
   \rightline{FSU-HEP-940311}\break
   \rightline{KEK-TH-392}\break
   \rightline{MAD/PH/825}\break
   \rightline{UH-511-785-94}
}}
\title{SUPERCOLLIDER SIGNATURES OF SUPERGRAVITY MODELS WITH YUKAWA
UNIFICATION}
\author{Howard Baer$^1$, Manuel Drees$^2$, Chung Kao$^1$, \\
Mihoko Nojiri$^3$ and Xerxes Tata$^4$}
\address{
$^1$Department of Physics,
Florida State University,
Tallahassee, FL 32306 USA
}
\address{
$^2$Department of Physics,
University of Wisconsin,
Madison, WI 53706 USA
}
\address{
$^3$Theory Group,
KEK, Oho 1-1,
Tsukuba,Ibaraki 305, Japan
}
\address{
$^4$Department of Physics and Astronomy,
University of Hawaii,
Honolulu, HI 96822 USA
}
\date{\today}
\maketitle
\begin{abstract}
We study the predictions of the simplest SU(5) grand unified model
within the framework of minimal supergravity, including constraints from
the radiative breaking of electroweak symmetry. As a consequence of the
unification of the $b$-quark and $\tau$-lepton Yukawa couplings,
the top quark mass is predicted to be close to its fixed point value. We
delineate the regions of the supergravity parameter space allowed by
constraints from the non-observation of proton decay and from the requirement
that the LSP does not overclose the universe. These constraints
lead to a definite pattern of sparticle masses:
the
feature unique to Yukawa unified models is that some of the third generation
squarks are much lighter than those of the first two generations.
Despite the fact that all sparticle masses and mixings are determined
by just four SUSY parameters at the GUT scale (in addition to $m_t$),
we find that the signals for sparticle production can vary substantially
over the allowed parameter space. We identify six representative scenarios
and study the signals from sparticle production at the LHC. We find
that by studying the signal in various channels, these scenarios may
be distinguished from one another, and also from usually studied ``minimal
models'' where squarks and sleptons are taken to be degenerate. In
particular, our studies allow us to infer that some third generation
squarks are lighter than other squarks---a feature that could provide
the first direct evidence of supergravity grand unification.
\end{abstract}
\medskip
%
%
%
\section{Introduction}

Supersymmetry (SUSY)\cite{REV} is increasingly being regarded
as the most compelling of the various extensions of the Standard Model proposed
to date. SUSY relates properties of bosons and fermions resulting in a new
level of synthesis. Supersymmetry also provides a technically natural
framework for incorporating the Higgs mechanism crucial for the breaking of
electroweak symmetry. On the experimental front, while the measurements of the
gauge couplings by the experiments at LEP are inconsistent with minimal models
of unification of gauge interactions, these measurements are in good agreement
with the predictions of the simplest supersymmetric grand unified
model\cite{UNIF}. As an added bonus, supersymmetry provides a natural candidate
for cosmological dark matter\cite{DM}.

Despite these many attractive attributes of supersymmetric theories, no
compelling model has as yet emerged. This could be due to the fact that
the dynamics of supersymmetry breaking is not yet understood. In
phenomenological analyses, supersymmetry breaking is incorporated by
the introduction of soft SUSY breaking terms which do not
jeopardize the stability
of the electroweak mass scale in the presence of radiative corrections
from very massive degrees of freedom not included in the SM (these
could be the particles of a grand unified theory (GUT), flavour physics
at intermediate scales or effects of quantum gravity at the Planck scale).
Even in the simplest low energy effective theories, the incorporation
of all possible such soft breaking terms\cite{GRIS} consistent with the known
low energy symmetries leads to a proliferation of
new parameters, greatly limiting the utility of this approach for SUSY
phenomenology.

This situation can be greatly ameliorated if the symmetries of the
interactions that lead to the breaking of supersymmetry can be guessed
at. For example, in a SUSY GUT, if SUSY breaking is assumed to occur
above the scale of GUT breaking, the soft SUSY breaking masses and couplings
of all particles in the complete multiplet are related. Supergravity (SUGRA)
models\cite{SUGRA}
provide a very attractive realization of this idea of utilizing
symmetries to restrict the supersymmetry breaking interactions. It is
assumed that SUSY
breaking in the effective low energy theory relevant for phenomenology
arises due to gravitational interactions, which being universal, allow for only
a few independent soft-supersymmetry breaking parameters (a common scalar mass,
a gaugino mass along with SUSY breaking trilinear, and possibly, bilinear SUSY
breaking scalar coupling) despite our ignorance of the details of the dynamics
of SUSY breaking. All this, of course, holds at the ultra-high scale of SUSY
breaking. In order to use the structure of the couplings thus obtained for
perturbative calculations at the 100 GeV energy scale relevant to experiment
today, we must sum the large logarithms arising from the disparity of these
scales. This is most conveniently done using the renormalization group
technique\cite{INOUE} whereby the soft breaking masses and couplings at the
scale of electroweak breaking\cite{SPECTRA,DN} may be obtained. When mixing
effects are negligible (as is the case for the first two generations of
sfermions) or absent (as for gluinos), these masses only differ slightly from
the physical (pole) masses of the sparticles\cite{POLE}. The real beauty of
this framework is that these same radiative corrections drive the desired
breakdown of electroweak symmetry, leaving colour and electromagnetic gauge
invariance intact. As a bonus, supergravity models lead to (approximate)
degeneracies of the first two generations of squarks which is
required\cite{CAMPBELL} by the non-observation of flavour changing neutral
currents in the K meson sector. It is interesting that the masses of $t$- and
$b$- squarks may be significantly different; this can lead to interesting
phenomenological consequences as we will see below.

While most phenomenological analyses of supersymmetry embody some of the ideas
of supergravity GUTs, the observation that the LEP data are consistent
with the predictions of SUSY SU(5) has led several groups\cite{SPECTRA,DN} to
systematically study the implications of the simplest SUGRA SU(5) model with
radiative symmetry
breaking constraints incorporated. It has been pointed out\cite{RAMOND}
that these
same data vitiate one of the earliest
successes of non-SUSY SU(5) GUT---the prediction of the
ratio $m_b/m_{\tau} \simeq 3$ from $b$-$\tau$ Yukawa coupling
unification. In contrast, the data are in agreement with the Yukawa
coupling unification in the SUGRA SU(5) framework provided that
the top quark mass is close to its infrared fixed
point\cite{MADGRP,POKORSKI,LANG}.

The purpose of this paper is to study the implications of the simplest
SU(5) supergravity model with radiative symmetry breaking constraints
incorporated. We will mainly focus on the detectability of the sparticle
signals at the Large Hadron Collider (LHC) and especially, on whether
these signals serve to distinguish this framework from the usually
studied ``minimal SUSY'' framework\cite{OTHER,BTW}.
Here, by simplest, we mean that we assume
the fewest number of particles
and the smallest number of soft-SUSY breaking parameters
at the scale $M_X$ of GUT symmetry breaking.
More specifically, we assume:
\begin{enumerate}
\item A minimal set of soft SUSY breaking parameters at the GUT scale
$M_X \simeq 2 \cdot 10^{16}$ GeV, i.e. a common scalar mass $m_0$, a common
gaugino mass $m_{1/2}$ and common trilinear ($A_0$) and bilinear ($B_0$)
soft breaking parameters.
We do not assume any relation between $A_0$ and $B_0$, though.
\item Minimal particle content at and below the scale $M_X$.
\end{enumerate}

We recognize that these assumptions are sensitive to the unknown physics at
high energy scales, and may well prove to be incorrect; for instance,
the assumption of a common gaugino mass might be invalid if there
are non-minimal kinetic terms for the gauge fields\cite{KIN}. Our
point, however, is that at present even this very restrictive model can
be claimed to be realistic in the sense it is compatible with all
experimental constraints both from particle physics as well as from
cosmology. Moreover, as we will see, the fact that the top quark Yukawa
couplings are driven to their fixed point values results in supercollider
signals that are characteristically different from previous studies\cite{BTW}.
Despite the fact that our analysis entails an
extrapolation of physics over a very large range of energy,
we believe that the simplicity of the model and the characteristic
predictions of the scenario provide sufficient motivation for our analysis.

In the next section, we discuss how the two assumptions together with
various observational constraints from collider and non-accelerator
experiments as well as from cosmological considerations further constrain
the model parameters in this already restrictive framework. We also
recapitulate how the sparticle spectrum and the SUSY parameters of
the low energy theory are obtained and present representative
examples of these. In section III, we study the general features
of the various scenarios obtained with our assumptions, and discuss
how these differ from those usually considered in phenomenological
analyses within the Minimal Supersymmetric Standard Model (MSSM) framework.
In the next section we briefly describe our simulation of sparticle
production at hadron colliders using
ISAJET 7.07\cite{ISAJET} and discuss various
improvements that we have recently incorporated into the program. In Section V,
we discuss various SUSY signals that might be expected at the Large
Hadron Collider (LHC) in the representative scenarios introduced above. We
show that an observable signal may be expected in various event topologies
including in single lepton and dilepton + $\eslt$ samples, which are
usually thought
to be obscured by
large SM backgrounds. Here, and in the next section,
we discuss how the simultaneous observation
of a signal in several channels can serve to differentiate the SUGRA SU(5)
from the MSSM framework, and also, to distinguish between the various
illustrative supergravity scenarios. We conclude with a summary of our results
and some general remarks.

\section {The Supergravity SU(5) Framework: Description of Constraints}

In this section, we first overview the relations between mass parameters at
the weak scale in the minimal SUGRA model. We then describe various
theoretical and experimental constraints that we impose, and show that despite
these, it is possible to find viable solutions even in our theoretically
restricted scenario.

At the high scale $M_X$, we start\cite{SPECTRA,DN} with the common gaugino
mass $m_{1/2}$, the common scalar mass $m_0$, the higgsino-mixing parameter
$\mu_0$, the soft SUSY breaking trilinear (bilinear) couplings $A_0$ ($B_0$),
together with the Yukawa couplings $h_t, \ h_b$ and $h_{\tau}$ (with
$h_b=h_{\tau}$ at scale $M_X$; see below). The absence of any intermediate
scales implies that all parameters at the weak scale can be determined
uniquely from the input parameters at $M_X$, by solving a set of
renormalization group equations (RGE) \cite{INOUE}. Of course, the above set of
parameters should reproduce the correct size of symmetry breaking, or $M_Z$,
$m_t$ (as a input), $m_b$ and $m_{\tau}$ at the weak scale; this leads to
constraints between the parameters at $M_X$. We do not assume any relationship
between $A_0$ and $B_0$. In practice we use the ratio of vacuum expectation
values $\tan \beta$ as an input parameter. Using the equations describing the
minimization of the Higgs potential at the weak scale then allows us to
express $B$ and $|\mu|$ at the weak scale in terms of $m_Z$ and $\tan \beta$
\cite{SPECTRA}; note that the renormalization group equations for $B$ and
$\mu$ decouple from all other RGE. The constraint of proper electroweak
symmetry breaking therefore reduces the number of free input parameters to
five, which we take to be $m_0, \ m_{1/2}, \ A_0, \ h_t$ and $\tan \beta$; in
addition, the sign of $\mu$ needs to be specified.

These considerations lead to simple relations between gaugino mass
parameters ($M_i$)
and first and second generation sfermion masses at the weak scale. The RG
running of the masses of strongly interacting particles implies large positive
corrections to squark and gluino mass parameters, while  corrections to the
electroweak interacting particles stay relatively small. Numerically,
$m_{\tilde{q}}^2 \simeq m_0^2+ 6 m^2_{1/2}$, and the soft SUSY-breaking
gaugino masses are related by $M_1 \simeq 0.4m_{1/2}$, $M_2
\simeq 0.8m_{1/2}$ and $M_3 \simeq 2.7m_{1/2}$.

The Higgs mass parameters and third generation squark masses receive
significant radiative
corrections from Yukawa interactions in addition to those from gauge
interactions. At $M_X$ our ansatz requires the two Higgs mass parameters
$m_{H_1}$ and $m_{H_2}$ to be equal: $m_{H_i}^2=m_0^2+ \mu_0^2$. As is well
known, this equality implies that the tree--level potential cannot
simultaneously lead to nonzero vacuum expectation values
and be bounded from below. Quantum
effects, i.e. the running of the mass parameters of the scalar potential, thus
become crucial for gauge symmetry breaking. Radiative $SU(2) \times U(1)$
symmetry breaking is in fact a natural prediction of this model, given the
large size of the top Yukawa coupling. More explicitly, $m_{H_2}^2$ receives
a negative radiative correction proportional to $h_t^2$ times some function of
soft SUSY breaking parameters when going towards the weak scale, due to its
interaction with the top quark. On the other hand $m_{H_1}^2$ is insensitive
to $h_t$, thereby lifting the equality between Higgs mass parameters. If $h_t$
is very large, $m_{H_2}^2$ becomes negative well above the weak scale,
resulting in too large a value of $m_Z$ or even an unbounded Higgs potential.
This can be compensated by choosing $|\mu|$ to be large. In other words,
$|\mu|$ has to be increased along with $h_t$ in order to maintain $M_Z$ at its
correct value.

We now turn to the constraints coming from our assumption of a minimal SUSY
$SU(5)$ GUT. The minimality of the $SU(5)$ Higgs sector immediately implies
the equality of the bottom and tau Yukawa couplings at the unification scale.
Several recent studies\cite{MADGRP,POKORSKI,LANG} have shown that this is only
compatible with the measured values of $b$ quark and $\tau$ lepton masses if
the top Yukawa coupling is large, i.e. close to its ``fixed point''
value\cite{FN1}. This conclusion persists even if the ratio $h_b/h_{\tau}$ is
varied about unity by 10\% to account for uncertainties due to threshold
effects. This immediately leads to the prediction
\begin{equation}
\label{sug1}
m_t(m_t) \simeq 190 \ {\rm GeV} \cdot \sin\beta ,
\end{equation}
where $m_t$($m_t$) is the running ($\overline{\rm MS}$) mass. Since we only
use 1--loop RGE everywhere, we simply interpreted eq.(\ref{sug1}) to mean
\begin{equation}
\label{sug2}
h_t (M_X) = 2.
\end{equation}
Taking the on--shell top quark mass $m_t \leq 190$ GeV\cite{BND}, we find
$\tan\beta \leq 2.5$ \cite{MADGRP}, in general agreement with the constraints
on it from nucleon decay\cite{DECAY} which we now turn to.

In SUSY GUTs, the exchange of higgsino triplets can lead to a dangerously
short lifetime of the proton. In minimal SUSY $SU(5)$ its exact value depends
on many parameters. In particular, the lifetime increases with increasing mass
of the higgsino triplet. Since the dimension--5 operators induced by higgsino
triplet exchange need to be ``dressed" by loops involving squarks and gauginos
(chargino loops usually being dominant), contributions from (s)quarks of
different generations in the loop add coherently. This opens the possibility
to increase the total lifetime, at the cost of some fine--tuning, by cancelling
contributions involving third generation (s)quarks against those involving
only the first two generations. The ``dressing loop function'' can be
decreased, and hence proton decay can be suppressed, by choosing scalars
(squarks and sleptons) to be considerably heavier than gauginos. For given
values of the other parameters the lower bound on the proton lifetime
therefore translates into an upper bound on the ratio $m_{1/2}/m_0^2$. In view
of the various uncertainties\cite{FN2} of this calculation we have
conservatively required\cite{DECAY},
\begin{equation}
\label{sug3}
m_0 \geq \min(300\ {\rm GeV},3 m_{1/2}).
\end{equation}
The coefficient 3 on the right hand side has been chosen so that the
contribution $m^2_0$ to $m^2_{\tilde{q}}$  described earlier dominates the one
$\propto m^2_{1/2}$, which implies $m_{\tilde q}\gg m_{1/2}$.

Finally, we require our model to be cosmologically acceptable. Because the
MSSM has a conserved R parity, the lightest supersymmetric particle (LSP) is
stable. Thus the LSP can be a good candidate for the missing cosmological mass
density, or Dark Matter (DM). Large $|\mu|$ means that the lightest
neutralino, which is also LSP in this scenario, is dominantly a gaugino.
The constraint eq.(\ref{sug3}) implies that LSP annihilation via the
$t-$channel exchange of sfermions is suppressed, leading to a dangerously
large relic abundance (or small age of the Universe) unless LSP annihilation
is enhanced by the proximity of an $s-$channel pole. This is not as unnatural
as it sounds, since recent studies\cite{AN2} have shown very large pole
enhancements even if the LSP mass is several GeV below half the mass of the
boson exchanged in the $s-$channel. Our large values of $\mu$ and $m_0$ and
the rather small value of $\tan\beta$ force most Higgs bosons to be heavy, the
exception, of course, being the lighter neutral scalar $H_{\ell}$. This is
because $m_{H_1}^2$
does not receive large negative radiative correction from Yukawa interactions
if $\tan\beta$ is small, while $m^2_{H_p}=m_{H_1}^2+m_{H_2}^2$ with
$m_{H_2}^2> -m_Z^2/2$ from the minimization of the Higgs potential. The
condition eq.(\ref{sug3}) then implies $m_{H_p} > 10 m_{\rm LSP}$. Therefore
only the $Z$ boson and the light Higgs boson can contribute significantly to
LSP annihilation.

In our computation of the Higgs sector, we have included one
loop corrections from top and bottom (the latter are small) Yukawa
interactions to both masses and trilinear Higgs interactions using the
effective potential\cite{POT}; note that these corrections are {\em not}
contained in the RGE discussed above. Despite the large top Yukawa coupling,
we find that $m_{H_{\ell}}$ is bounded from above by about 110 GeV for the
values
of $\tan\beta$ we are considering. Moreover, $H_{\ell}$ is always SM--like in
this
scenario, due to the large masses of the other Higgs bosons. Hence the LEP
bound of 62.5 GeV applies for it as well. This then leads to bounds on the
mass of a cosmologically acceptable LSP, and hence on the gaugino mass:
\begin{equation}
\label{sug4}
60 \ {\rm GeV} \leq |m_{1/2}| \leq 130 \ {\rm GeV}.
\end{equation}

Finally, we impose the limit
\begin{equation}
\label{sug5}
m_0 \leq 1 \ {\rm TeV}
\end{equation}
in order to avoid excessive fine tuning.

We should mention a technical complication that arises from the constraint
(\ref{sug3}) as a result of which squarks are substantially heavier than
gluinos. It has been shown\cite{POLE} that the pole gluino mass receives
additional corrections proportional to $\log (m_{\tq}/m_{\tg})$. Since one of
the features of these scenarios, as we will see, is a rather light gluino,
these extra corrections -- which can be as large as 30\% -- can be
significant. We have incorporated these into our program using the formulae in
Ref.\cite{POLE}.

Our apparently innocuous assumption of minimal particle content at the GUT
scale led to the various constraints on model parameters as we have just
discussed. Most analyses require {\em either} the equality of $b$ and $\tau$
Yukawa couplings at $M_X$, {\em or} an acceptably long lifetime for the
proton, but not both. It seems to us that these two constraints are really on
very similar theoretical footing in that they can both be weakened or even
avoided completely if one allowed a more complicated Higgs sector and/or a
larger GUT group\cite{FN3}. On the other hand, we do not try to find
correlations between the value of the strong coupling constant and the overall
size of SUSY breaking masses. Given the present (mostly theoretical) error of
$\alpha_s (M_Z)$, and in view of further uncertainties at the high scale we do
not consider any ``bounds'' on sparticle masses derived in this fashion to be
reliable at present. We now turn to a discussion of the characteristics of the
fixed point scenario outlined above, and give some illustrations of the
spectra and the sparticle mixing patterns that may be anticipated in this
class of models.

\section {The Supergravity SU(5) Framework: Sparticle Masses and Mixing
Patterns}

The characteristic feature of the fixed point scenario of the last section
is the large top Yukawa coupling, eq. (\ref{sug1}). This, as we have
seen, leads to $|\mu |\geq m_0$, assuming the
radiative breaking mechanism
of electroweak symmetry.
Together with eq. (\ref{sug3}), this then implies
that the lighter (heavier) charginos and neutralinos are gaugino- (higgsino-)
like in character. The limit eq. (\ref{sug4}) implies that gluinos
cannot be much heavier than about 400 GeV and may even be light
enough to be discovered at the Tevatron.
We have also seen that the pseudoscalar ($H_p$), the
charged ($H^\pm$) and the heavier neutral scalar ($H_h$) Higgs bosons all
tend to be heavy, so that these will not be produced via cascade decays
of gluinos\cite{HIGCAS}. Finally, we make the obvious remark,
$m_{\tq}> m_{\tg}$ so that the squarks of the first two families
dominantly decay to gluinos.

Another important consequence of the large top family Yukawa interactions
is the sizable reduction of the SUSY--breaking masses of
scalar top quarks, and further, the mass squared of $\tilde{t_R}$ is reduced
twice
as much as that of $\tilde{t_L}$.
By $SU(2)$ invariance this means that the mass of left--handed scalar bottom
quarks will also be reduced compared to squarks of the first two generations.
The top squark masses are further altered by squark mixing, which depresses
the lighter state even further.
This non--degeneracy of third generation squarks, which is one of the
characteristic
features of minimal SUGRA models with radiative gauge symmetry breaking is
accentuated by the large value of $h_t$ in the model.

In spite of the rather strong constraints on parameter space, we found that
there is still considerable room for variation in the expected phenomenology.
While condition (\ref{sug3})
implies that masses of the first and second generation squarks are well above
the gluino mass no such statement can be made for the lighter stop
eigenstate $\tilde{t_1}$. In particular, it is possible that the two body decay
$\tilde{g} \to t + \tilde{t_1}$ is allowed\cite{FN4}, in which case it will
have a branching ratio near 100\%, being the only two--body decay mode of
$\tilde{g}$. We remind the reader that it has been shown in Ref.\cite{BDGGT}
that signals from a relatively light $t$-squark may well have escaped
detection at the Fermilab Tevatron, although with the accumulation
of a large data sample, its detection may be possible\cite{BST}.

Despite the various constraints that we have imposed
it is impossible to do an exhaustive scan of parameter space. Moreover, for
the purposes of this study, the main question is not whether it is possible
to detect supersymmetry at the LHC (for the light spectrum we have
this has already been answered affirmatively\cite{BTW}), but whether
it is possible to distinguish the present class of models from the
usually studied MSSM, and further, whether it is possible to distinguish
the various scenarios (discussed shortly) from one another. Toward
this end, we have attempted to choose model parameters (respecting
the constraints discussed in Sec. II) that accentuate the qualitative
differences in the signals.

At hadron colliders, squarks and gluinos are the most copiously produced
particles. Since, within our framework, the first two
generations of squarks dominantly decay
to gluinos, we expect that the gluino decay patterns fix the broad
features of the phenomenology. This immediately leads us to consider
two classes of scenarios: (A) where the decay $\tilde{g} \to t + \tilde{t_1}$
is allowed, and (B) where it is forbidden. Case A naturally subdivides
into two cases determined by the decay pattern of the lighter $t$-squark.
In case A1, we assume that the decay $\tilde{t_1}\to b\tw_1$ is accessible;
since this is the only tree-level
two body decay, its branching fraction is essentially 100\%.
In case A2, we assume that the decay $\tilde{t_1}\to b\tw_1$ is kinematically
inaccessible; in this case the stop decays\cite{HK} via one loop diagrams
into a charm jet and an LSP: $\tt_1\to c\tz_1$.

Turning to the heavier stop, case B, we have attempted to diversify the
phenomenology by choosing combinations of $m_0$ and $m_{1/2}$ at their
maximum and minimum values discussed in Sec. II. This leads to four
cases B1--B4.
\begin{itemize}
\item B1: $m_{1/2}$ and $m_{0}$ small,
\item B2: $m_{1/2}$ small; $m_{0}$ large,
\item B3: $m_{1/2}$ large; $m_{0}$ small,
\item B4: $m_{1/2}$ and $m_{0}$ large.
\end{itemize}
Notice that for the case B1 where both squarks and
gluinos are at their minimum phenomenologically acceptable
values, the gluino may even be
accessible at the Tevatron. Our point in analysing this case is more
to illustrate the diversity of phenomena that could result even
in this restricted framework, and to study the range of parameters
over which distinguishable signals may be possible at the LHC.

The input parameters, the resulting spectra and the values of $\mu$ at the weak
scale in these six cases are shown in Table I. Several features are
worth pointing out.

\begin{itemize}
\item We see that the gluinos are relatively light. As noted above, gluino
signals in case B1 (and, possibly, B2) may even be observable at the Tevatron.
\item It is striking to see that
in case A2, $m_{\tt_1}$ is only 84 GeV; a significant
number of $\tt_1$-pair events should, therefore, be present in the existing
data sample at the Tevatron. As pointed out in Ref.\cite{BDGGT} such a
light stop squark may well have escaped detection at the Tevatron especially
since the LSP is rather heavy in this case. It should, however, be remembered
that both the experiments at the Tevatron have since accumulated at least
15-20 $pb^{-1}$, so that the detectability of the stop at the Tevatron
merits a reevaluation\cite{BST}. It would indeed be interesting to study
whether
stop squark signals in cases A and B2 might be detectable at the Tevatron.
\item The lighter chargino and the second lightest neutralino in B1 and B2
should be kinematically accessible at LEP II. For the other cases in Table I,
their accessibility will obviously depend on the energy that LEP II is
finally able to attain. Since, as remarked above, both $\tz_1$ and $\tz_2$
are dominantly gaugino-like (and selectrons are heavy), their production at LEP
II will be suppressed. For cases B1 and B2, the trilepton signal from
$\tw_1\tz_2$ production at the Tevatron should be present at observable rates.
In the other cases illustrated, the trilepton cross section (after experimental
cuts) is $\sim$1 $fb$\cite{BKT}, and hence, is unlikely to be observable.
This is because we have $m_{\tl} \simeq m_{\tq}$ so that, unlike as in no-scale
models, the leptonic branching ratios are not dynamically enhanced.
\item The
light Higgs boson $H_{\ell}$ is essentially the SM Higgs boson. It should,
therefore, certainly be detectable at LEP II in case B1, while the
detectability in the other cases will be sensitive to the machine energy that
is ultimately achieved.
\item  We note that the top quark is necessarily within the reach
of the Tevatron, and further, its decays are as given by the SM except
in case A2 where the decay $t\to \tt_1\tz_1$ has a branching fraction
of $< 10$\%.
\item Finally, we see that as noted above, $m_{\tb_L}$ is indeed considerably
smaller than the mass of the first two generations of squarks. This
implies that three body gluino decays mediated by $\tb_L$ exchange are
enhanced relative to those mediated by other squarks. As a result, gluino
events in cases B may be expected to have a higher multiplicity of $B$ mesons
than would be expected if all squarks are taken to be degenerate as in
most MSSM studies. It is amusing to note that $\tg\tg$ events may
be even richer in $B$'s for cases A, since then, decays of $t$ quarks from
$\tg \to \tt_1 t$ as
well as the decays $\tt_1\to b\tw_1$ (in case A1) lead to $B$ meson production.
We also note that because $\tb_L$ is relatively light, $\tz_2 \to
b\bar{b}\tz_1$
decays can also be considerably enhanced relative to $\tz_2 \to d\bar{d}\tz_1$
decays. We thus see that the presence of light third generation squarks
leads to SUSY events that should be rich in $B$ mesons. Such events
may also occur within the MSSM framework where cascade decays of gluinos
to Higgs bosons can result in events with a high multiplicity of
$b$ quarks\cite{BBTW}. However, these cascade decays can occur only
when $m_{\tz_2}-m_{\tz_1}$ exceeds the lighter Higgs boson mass which, for the
SM Higgs bosons, is bounded below by 63.5 GeV\cite{BOUND}. Thus, within
the MSSM framework, these events occur only when gluinos are rather heavy.
In other words, the detection of $B$-rich gluino events with event kinematics
corresponding to gluino masses smaller than 400 GeV could signal a light
third generation of squarks.
\end{itemize}

Next, we briefly describe the improvements that we have made in
ISAJET 7.07\cite{ISAJET} to simulate supersymmetry events at hadron colliders
that we use to compute event rates, topologies and kinematic distributions
for the six cases listed in Table I.

\section {Simulation of Supersymmetry at Hadron Colliders}

The main production processes and decay modes of all sparticles as given
by the minimal supersymmetric model have
been incorporated into ISAJET\cite{ISAJET}
which we use for all our simulations. In order to allow for
the possibility that the third generation of squarks may be substantially
lighter than the first two generations, which is the distinguishing
characteristic of the scenario, we have had to make several
modifications to the code. In ISAJET 7.07,

\begin{enumerate}
\item we have included the production and decays of $t$-squarks.

\item We have modified the input parameters to the program.
In addition to the SM parameters (including $m_t$) the current low energy
inputs in terms of which sparticle masses and decay patterns are computed
are ($m_{\tg}$, $m_{\tq}$, $m_{\tl_L}$, $m_{\tl_R}$, $m_{\tnu}$, $\tan\beta$,
$\mu$, $m_{H_p}$, $m_{\tt_L}$, $m_{\tt_R}$, $A_t$, $m_{\tb_R}$, $A_b$).
The last five entries are new modifications
and allow for independent mass and mixing
pattern for the third squark generation. The stop and sbottom masses
are the respective soft SUSY-breaking mass parameters and not the
physical squark masses. Since $\tb_L$ and $\tt_L$ belong to the
same SU(2) doublet, $m_{\tb_L}=m_{\tt_L}$. We have neglected D-term
contributions that break the degeneracy between the L- and R-type
squarks of the first two generations but have retained
these in the computation of slepton masses---this is the only
reason for the difference in the $\tl_L$ and $\tnu_{\ell_L}$ masses.
For heavy squarks, this D-term
splitting is generally small, and so, makes little difference in
the computations of squark production cross sections or gluino,
chargino and neutralino decays mediated by virtual squarks. This
splitting can be very important
when squarks and gluinos are close in mass,
since then some of the two-body gluino decay modes may become
accessible. In the present study, this
is never the case because of the constraint (\ref{sug3}).

\item The input gluino mass is the physical pole mass\cite{POLE} as opposed
to the running $\overline{MS}$ mass that enters the
gaugino-Higgsino mass matrices. In all the cases in table I, the
pole mass is somewhat larger than the $\overline{MS}$ mass, primarily
because the squarks are heavier than gluinos. For Case B2, where
the gluino is light and squarks are at their maximal value, this
difference is almost 50 GeV.

\item In the computation of the Higgs boson masses and couplings, we have, as
already mentioned, incorporated radiative corrections from both top and bottom
Yukawa interactions, including mixing between
third generation squarks. We have not included effects of virtual gauge boson
and gaugino loops, but these are known to be small\cite{VIENNA}.

\item We have supplemented the various sparticle decay formulae to
account for non-degenerate third generation sparticles. This can be
especially important for gluino three body decays, as well as
the corresponding neutralino decays. We have also included
contributions to three body $\tz_2$
decays
mediated by Higgs bosons;
these can be important for large values of $\mu$ and $m_{\tq}$
for which the contributions from both $Z$ and squark mediated
decays are dynamically suppressed. As a result, $\tz_2$ decays to
heavy quarks and leptons may be enhanced.
\end{enumerate}

For every set of input parameters, the routine ISASUSY calculates all
sparticle masses along with branching fractions for all their allowed
decay modes. Of course, the input parameters are correlated by the
minimal SU(5) supergravity constraints discussed above.
We then use ISAJET to generate supersymmetric particle pairs
in $pp$ collisions. In each
run, ISAJET is set up to simultaneously generate all allowed sparticle pairs
in numbers proportional to the corresponding pair
production cross section. We use the EHLQ Set I structure
functions\cite{EHLQ} for our computations of cross sections at the LHC. The
produced sparticles are then decayed via various cascades with
branching fractions as obtained from ISASUSY into final states
involving quarks, leptons, gluons and photons. Radiation off initial and
final state partons as well as heavy flavour fragmentation is included in
ISAJET. Daughter quarks and gluons are hadronized, and unstable
particles decayed into the final state. Finally, underlying soft scattering
and hadronization of beam remnants is modelled in our simulation.

For each of the six cases in Table I, we use ISAJET  to generate a
sample of 50K SUSY events by $pp$ collisions at $\sqrt{s}=14$ TeV.
The relative contributions of the different SUSY processes to this
event sample is shown in Table II. We see that for the light stop
cases A1 and A2, the production of stops dominates, and furthermore,
that the stops have a significant production rate even in most
of the cases B. The production of other flavours of squarks
is suppressed relative to that of gluinos primarily because
of Eq.(\ref{sug3}). We have also shown the branching fractions
for the main decay modes of sparticles that are most abundantly
produced (either directly or via cascades) at the LHC. In cases
A1 and A2, the gluino essentially always decays to $t$-squarks
whereas in cases B, $\tw_1$ and $\tz_2$ decays are favoured\cite{BBKT}.
We also see that gluino decays to $b$ quarks are enhanced relative
to the corresponding decays to $d$ quarks. This is because $\tb_L$
is lighter than the other down type squarks. This enhancement
is small for cases B1 and B2 because the $\tz_1b\tb_L$ coupling
is suppressed by accidental cancellations --- these cancellations
do not occur for gluino decays to $\tz_2$. Notice also that the decays
$\tz_2\to b\bar{b}\tz_1$ are also enhanced relative to
$\tz_2\to d\bar{d}\tz_1$ decays, in part for the same reason, but also because
of $H_{\ell}$ exchange diagrams. Finally,
we have also shown the branching fraction for the leptonic decays of
$\tw_1$ and $\tz_2$ which enter the computation of the multilepton
signals. Notice that the neutralino branching fraction exhibits substantial
variation while that for the chargino is essentially determined
by the W branching fraction except in Case A2 where the chargino
decays via the two body $\tt_1 b$ mode.

In order to simulate
experimental conditions at the LHC,
we impose cuts described in the next section and then classify the
signals according to their lepton content as
in Ref.\cite{BTW}. Since the efficiency for SUSY events to pass these
cuts depends strongly on the masses and decay patterns of the
produced sparticles,
the contribution of the various processes {\it after cuts} is, in general,
quite different from the relative rates in Table II.
To compare the features of sparticle production in
the supergravity SU(5) model under study, with the corresponding production
in ``minimal SUSY'' models usually\cite{BTW,OTHER} used to study SUSY
signatures, we repeat our simulation for four illustrative cases. Here, we take
all squarks and sleptons to be degenerate, and fix the gluino mass at 300 GeV,
a value similar to $m_{\tg}$ in the six cases in Table I. Since the
signals are known to be sensitive to $m_{\tq}/m_{\tg}$, we illustrate
our results for $m_{\tq}=m_{\tg}+20$ GeV and $m_{\tq}=2m_{\tg}$. To vary
the electroweak gaugino content we study two values of $\mu$. Specifically,
the parameters for these four cases are:
\begin{itemize}
\item BTW1: $m_{\tg}=300$ GeV, $m_{\tq}=320$ GeV, $\mu=-150$ GeV,
\item BTW2: $m_{\tg}=300$ GeV, $m_{\tq}=600$ GeV, $\mu=-150$ GeV,
\item BTW3: $m_{\tg}=300$ GeV, $m_{\tq}=320$ GeV, $\mu=-500$ GeV,
\item BTW4: $m_{\tg}=300$ GeV, $m_{\tq}=600$ GeV, $\mu=-500$ GeV.
\end{itemize}

For comparison purposes, and in keeping with parameters of Ref. \cite{BTW},
we have fixed $\tan\beta=2$, again a value similar
to the cases in Table I, and chosen $m_t=140$ GeV, $m_{H_p}=500$ GeV, with
the $A$-parameters set to zero. A discussion of our simulation along with
the various results forms the subject of the next section.

\section{SUPERGRAVITY SU(5) SIGNALS AT THE LHC}

As emphasized in Ref.\cite{BTW}, several interesting signals should
simultaneously be present if any new physics signal at the LHC is to be
attributed to squark or gluino production. These signals include the
canonical $\eslt$ events, the like-sign dilepton events and multilepton
events. These leptonic events should contain substantial jet activity.
The production of charginos or neutralinos in association with
a squark or gluino yields the same event topologies, while
chargino-neutralino production could, in addition, lead to multilepton
events free of jet activity. In this section, we study the rates and
distributions for these various event topologies for the six supergravity
cases in Table I and compare these with the ``minimal supersymmetry'' test
cases listed above.

We use the toy calorimeter simulation package ISAPLT to model detector effects.
We simulate calorimetry with cell size
$\Delta\eta\times\Delta\phi =0.1\times 0.1$, which extends between
$-5<\eta <5$ in pseudorapidity. We take electromagnetic
energy resolution to be $10\% /\sqrt{E_T}\ \oplus\ 0.01$, while hadronic
resolution is $50\% /\sqrt{E_T}\ \oplus\ 0.03$ for $|\eta | <3$, and
$100\% /\sqrt{E_T}\ \oplus \ 0.07$ for $3<|\eta |<5$, where $\oplus$
denotes addition in quadrature.
Jets are coalesced
within cones of $R=\sqrt{\Delta\eta^2 +\Delta\phi^2} =0.7$ using
the ISAJET routine GETJET. Hadronic clusters with $E_T>50$ GeV
are labelled as jets.
Muons and electrons are classified as isolated if they have $p_T>20$ GeV,
$|\eta (\ell )|<2.5$,
and the visible activity within a cone of $R =0.3$ about the lepton
direction is less than $E_T({\rm cone})=5$ GeV.
Finally, we assume (perhaps optimistically) that it is possible to identify
$b$-jets with $p_T>$20 GeV within $|\eta|<2$ with an efficiency of
40\% via micro-vertex tagging.

Following Ref. \cite{BTW}, we group the various signal categories
according to the number of isolated leptons, and further, require
$\eslt > 100$ GeV for the SUSY signal. The various signals are,

({\it i}) the $\eslt$ signal consisting of $0\ell +n_j\ jets+\eslt$ events.
We require $\eslt > 150$ GeV, $n_j \geq 4$ and the transverse sphericity
$S_T > 0.2$;

({\it ii}) the single lepton signal consisting of events
with one and only one isolated lepton ($1\ell +jets+\eslt$);

({\it iii}) events with two isolated leptons ($2\ell +jets+\eslt$), which
can subdivide into opposite sign dileptons (OS) and same sign dileptons (SS).

({\it iv}) events with three isolated leptons ($3\ell +jets+\eslt$);

({\it v}) and events with four isolated leptons ($4\ell +jets+\eslt$).

Table III gives an overview of the cross sections for these event topologies
for the six SUGRA cases in Table I along with the cross sections for the major
SM backgrounds. Also shown are the corresponding cross sections for the four
``MSSM'' cases introduced in Sec. IV. The cross sections shown here
include the various cuts mentioned above, as well as additional
cuts (discussed below) for the $\eslt$, the $1\ell$ and the OS dilepton
signals which are known to have large SM backgrounds.
As we will see, these additional cuts are essential to suppress
the SM background, particularly to the $1\ell$ and $\ell^+\ell^-$ signals
to acceptable levels. We see from Table III that for all but
the $1\ell$ signal in Case B4, the signal to background ratio exceeds unity,
and is frequently considerably larger. Furthermore, we see that the cross
sections in the $\eslt$, $1\ell$ and dilepton channels range from 1-100 $pb$
so that $10^4$-$10^6$ signal events may be expected annually at the LHC
even for a luminosity of $10^4$ pb$^{-1}$/year. We also see that except
for case A2, several thousand trilepton events may be expected
every year at the LHC. We have been unable to identify any significant SM
physics
background to this signal. Finally, we note that while there is some
variation of the signals for the different scenarios in Table III,
these cross sections alone do not enable us to unambiguously distinguish
between
all the various cases, or for that matter even between the SUGRA and
the ``MSSM'' cases. This requires a more detailed study of the various
signals which we now turn to.

\subsection{$\eslt$ signal}

This is generally regarded as the canonical signal for supersymmetry.
Backgrounds in this channel can be very dependent on detector details,
especially forward calorimetry, and realistic BG estimates require a
detailed calorimeter simulation. Nevertheless, we estimate backgrounds
from several major SM sources using our toy calorimeter and ISAJET:
$t\bar t$ pair production, for $m_t=160$ GeV, $W+jets$ production, $Z+jets$
production, and heavy flavor ($b\bar b$ and $c\bar c$) production. Considerable
background rejection with relatively little cost to signal can be gained by
requiring, in addition,
\begin{itemize}
\item $\Sigma |E_T|>700$ GeV, where the sum is over all calorimeter cells.
\end{itemize}
The impact of this cut on the signal and the various backgrounds
is shown
in the $\eslt$ column of Table IV. The values (inside) outside the
parenthesis indicate the signal cross section (before) after this
additional cut. While this cut is not essential in that the signal
to background ratio substantially exceeds unity even before the cut,
we see that implementing the cut results in a significant improvement
of this ratio with only a modest loss of signal. The largest $\eslt$ cross
sections
come from the low $m_{1/2}$ (light gluino) cases, where the total production
cross sections are largest. In order to facilitate distinction between
the different scenarios in Table III, we exhibit
various characteristics of the $\eslt$ events in Table V.
In columns 3-5, we list the fraction of events with
jet multiplicities $n_j =4-5$, $n_j=6-7$ and $n_j\ge 8$. The most noticeable
difference occurs in the $n_j\ge 8$ channel, where we see several SUGRA
cases yield large fractions of events with very high jet multiplicity.
This is especially true of cases A1, A2 and B4. For instance, in A1,
many of the events come from gluino pair production, where
$\tg\to t\bar{\tt_1}\to bqq'+bqq'\tz_1$, which can yield up to
twelve `parton level' jets. Of course, it is very difficult for the SM
background processes to yield such high jet multiplicities.

We also list in Table V the fraction of events containing identifiable
displaced $b$-vertices, $n_b$.
An important point is that supersymmetry events
often have high multiplicity of
$B$ hadrons\cite{BTW,BBTW} in the central region. This is even more
so in the SUGRA cases we have generated. We see that in the SUGRA cases,
from $25\% -65\%$ of the $\eslt$ events have at least one identifiable $B$
vertex, compared to $15\% -25\%$ for the BTW cases (of course, larger
B multiplicities are also possible in the BTW cases when cascade
decays of gluinos into Higgs bosons become operative. However,
this occurs for considerably larger values of $m_{\tg}$.). Also, the SUGRA
cases
can have large fractions of events with $n_b\ge 2$ and $n_b\ge 3$. There are
several reasons for this:
\begin{itemize}
\item in case A1 with the richest $B$ content, each gluino in gluino pair
events decays via $\tg\to t\tt_1$, which automatically yields at
least four $B$'s per event since the $t$-quark as well as the $\tt_1$
squark decay into a $b$,
\item the fact that $\tb_L$ is generally lighter than the first
two generation squarks means that three body gluino decay via a virtual $\tb_L$
is enhanced, yielding a large branching fraction for
$\tg\to b\bar{b}\tz_i$\cite{DN},
\item neutralino three body decays such as $\tz_2\to b\bar{b}\tz_1$ are
frequently enhanced, either because of the smaller $\tb_L$ mass, or due
to the importance of Higgs boson mediated diagrams discussed in Sec. IV.
\end{itemize}
Of course, the exact $b$ multiplicity will depend on the tagging
efficiency that is ultimately attained.
The ability to tag $B$ vertices may also be very helpful in
obtaining an enriched SUSY sample, and thus aid in SUSY particle mass
estimates.

We also show in Table V the mean summed scalar $E_T$ and mean $\eslt$ in
$\eslt +jets$ events. Use has already been made of $\Sigma |E_T|$ to
enhance SUSY signal against background. The scalar $E_T$ distribution
is somewhat sensitive to
the magnitude of the SUSY particle masses produced in the
subprocesses although this is not clear from just the mean values
listed in Table V. We also remark that the differences between
the various cases are somewhat reduced by the hard $\Sigma |E_T|$ cut.
We show in Fig. 1 the $p_T$ distribution of the fastest jet
in $\eslt$ events, for four SUGRA cases, and the sum of backgrounds. The SUGRA
cases have $p_T$ which peaks around $\sim 200-250$ GeV, with a broad tail
extending to higher $p_T$'s, unlike SM background. Since squarks of the
same mass are expected to lead to harder jets than gluinos, it is natural
to ask whether squark production shows up as a shoulder in this distribution.
A glance at Fig. 1 shows that this will be difficult. We remark though
that the hard cuts that we have imposed to enhance the signal selects
out events with similar kinematics and so reduces the differences
between the various signals.

\subsection{Single isolated lepton events}

Little effort has gone into searches for SUSY in the single lepton channel,
due in part to large backgrounds expected from $W+jets$ production
and $t\bar t$ production. Within our framework, the production rate
of (the rather light) gluinos is truly enormous at the LHC so that
one may hope to be able to make rather stringent cuts to eliminate
the backgrounds, leaving behind an observable signal.
A glance at the cross sections in parenthesis
listed in the $1\ell$ column of Table IV shows that
the single lepton cross section is smaller than the SM background
except in case B1. To facilitate further separation between
the SUSY signal and the SM background, we show in Fig. 2
{\it a}) the $\Sigma |E_T|$ and {\it b}) transverse mass
$m_T(\ell ,\eslt )$ distributions for four SUGRA cases A1, A2, B1 and B4,
along with the summed background distribution. The scalar $E_T$ distribution
is substantially harder than SM background.
Since both the top and the $W$ backgrounds dominantly contain one
leptonically decaying $W$ boson, the transverse mass
distribution for the background rises to a Jacobian peak at $m_T=M_W$ while
that for the signal is relatively smooth. Hence, to
maximize signal against background, we further require:
\begin{itemize}
\item $\Sigma |E_T|>700$ GeV, and
\item $m_T(\ell,\eslt )<60$ GeV, or $m_T(\ell,\eslt )>100$ GeV.
\end{itemize}
The resultant cross sections are shown in Tables III and IV, in the $1\ell$
column. The total SM background is $\sim 12\ pb$, while signal
ranges between $10-77\ pb$. In order to get some idea of the hardness
of the leptons in the $1\ell$ sample,
the $p_T(\ell )$ distribution after all the above
cuts is shown in Fig. 2{\it c}.
In case B1, where single leptons are apt to come from
light chargino decay, a distortion of the low $p_T(\ell )$ distribution
can occur; for other cases, where leptons come from $t$-quark
decay, the distribution more nearly tracks the shape of the
background.

Various characteristics of single lepton events are detailed in
Table VI. The general features are similar to those of Table V.
The jet multiplicity is again large, especially for events
with larger gluino masses, while jet multiplicities for SM backgrounds
are substantially smaller. Furthermore, the SUSY signal events,
especially for cases A1, A2 and B4, are characterized by large
fractions with visible displaced $B$ vertices. The SUGRA cases, except
for case B2, all have $n_b\ge 2$ substantially
larger than the corresponding BTW
values, and hence could yield evidence for a SUGRA-type of sparticle
spectrum, where some of the third generation squarks are far lighter
than squarks of the first two generations.

\subsection{Opposite-sign (OS) isolated dilepton events}

As for the case of single lepton events, searches
for SUSY in the isolated opposite sign dilepton channel have
rarely been considered due to a large expected background from
$t\bar t$ events. Within our framework, however, we see from
the cross sections in parenthesis in the OS column in Table IV
that, with just the canonical cuts discussed above, the dilepton signal
exceeds that background in the favourable cases A1, B1 and B2. Furthermore,
it is only between a factor 2-6 smaller than the background in the other
cases. In Fig. 3{\it a}
we plot the $\Sigma |E_T|$ distribution, for which the signal gives a
substantially harder spectrum than background. We see that the additional
requirement,
\begin{itemize}
\item $\Sigma |E_T|>700$ GeV,
\end{itemize}
reduces the background by a factor of seven, with only modest
loss of signal, particularly for the cases A2, B3 and B4 where
the signal was considerably below background.
The cross sections after all cuts are shown in Tables III and IV, in the OS
columns. Signal exceeds background for cases A1, B1 and B2, and is
comparable in the other SUGRA cases.
The $p_T(\ell )$ distribution for fast and slow leptons  after the above
cuts is shown in Fig. 3{\it b} and 3{\it c}. The overall hardness of these
distributions depends in part on the SUSY sources of the dileptons. In
cases A1 and A2, many of the leptons are quite hard, coming from top quark
decay,
while in cases B2 and B4 the leptons frequently come from three body
chargino and neutralino decay, and exhibit a softer spectrum. For OS dileptons
produced in gluino and squark events, either
the leptons each separately come from the two SUSY particles produced in the
hard scattering, or the two leptons can come from a single SUSY particle
cascade, such as $\tg\to\tz_2\to \ell\bar{\ell}\tz_1$. In the former
case, the dilepton pair is expected to have a large opening angle in
the transverse plane, whereas in the latter case, the opening angle is expected
to be small. This is borne out in the behavior of Fig. 4. In cases B2 and
to a smaller extent in case B4 (cases B1 and B3 are similar to B1 and B4,
respectively),
where there is substantial neutralino production via cascade decays,
the dileptons come with a small opening angle. In case A1 the dileptons
are somewhat  peaked at large opening angle, though the peak
is not as pronounced as in cases B2 and B4. This is because the OS dileptons
can come from either the same gluino (from the leptonic decay of its
$t$ and $\tt_1$ daughters) or from different gluinos and the cuts
tend to favour leptons from top decays to those from the
decays of stop.

We show OS lepton event characteristics in Table VII.
The SUGRA cases again show large jet multiplicity, and large $B$
hadron multiplicity; the latter quantity can again serve to distinguish
SUGRA events from the MSSM BTW cases.

Just as with dilepton opening angle, the SUSY source of OS dileptons
is correlated with the dilepton flavor asymmetry,
defined as
\begin{equation}
\label{sug6}
A_f={{N(e\bar{e})+N(\mu\bar{\mu})-N(e\bar{\mu})-N(\bar{e}\mu )}\over
{N(e\bar{e})+N(\mu\bar{\mu})+N(e\bar{\mu})+N(\bar{e}\mu )}}
\end{equation}
SS dilepton events are expected to have $A_f\sim 0$ (replacing $\bar{\ell}$
with
$\ell$), reflecting
uncorrelated sources of the leptons; however, OS dileptons may be expected to
have a large positive $A_f$ if many of the OS dileptons come from
{\it e.g.} $\tz_2\to\ell\bar{\ell}\tz_1$ decay. The asymmetry is listed
in the last column of Table VII. Indeed, cases B1 and B2 have large $A_f$,
while A1 and A2 have $A_f$ consistent with zero. Cases B3 and B4 show
intermediate values of $A_f$. Notice that the flavour asymmetry qualitatively
tracks the azimuthal angle peaking illustrated in Fig. 4.

\subsection{Same-sign (SS) isolated dilepton events}

The same-sign isolated dilepton signature has been
recognized\cite{BGH,BTW} for several
years now as a particularly clean signal for SUSY, owing to tiny SM
backgrounds. It is indicative of the Majorana nature of the gluino; in a gluino
pair production event, both gluinos may decay to charginos of the same sign so
that subsequent chargino leptonic decay leads to the same-sign dilepton
signature.
The cross sections after the cuts discussed above are listed in Table III,
under the $SS$ column.
The SUGRA cases yield cross sections between 0.4-3.7 $pb$ at the LHC,
which would yield 4000-37,000 events for an integrated luminosity of
$10^4$ $pb^{-1}$ per year. Backgrounds are expected to come
from $t\bar t$ production, where a lepton from $b$ decay is
somehow recorded as isolated and from $W^\pm W^\pm$ production.
Our simulation of 300K $t\bar t(160)$
events gives a background of 0.01 $pb$, well below signal whereas the
corresponding background from same sign $WW$ production has been shown
to be\cite{DV,BTW} $\simeq 0.035$ $pb$. We thus expect that this
sample is essentially free of {\it physics} backgrounds.

Each lepton in a SS dilepton event usually originates from the cascade decay of
different particles produced in the hard scattering subprocess. Thus, the
transverse SS dilepton opening angle $\phi(\ell\ell ')$ distribution
should be maximal for
large opening angles. This is shown by the $\phi$ distribution plotted
in Fig. 4. This behavior, in comparison to that of OS dileptons,
reflects the source of the dilepton parentage, and hence may be useful in
ultimate event reconstruction. Furthermore, if SS and OS dileptons had the same
source, for instance as in $\tg\tg$ events, then one would expect roughly
equal rates for SS and OS dilepton production. In Table III, we present
the ratio of OS/SS event rates, where the same cuts have been made for
each topology, {\it i.e.} the $\Sigma |E_T|$ cut has been removed for OS
dileptons. We see in case A2, where dileptons come mainly from $t$ decay
in gluino pair events with $\tg\to t\tt_1$, that this is indeed the case.
In case A1, leptonic decays of the top and stop daughters of the same gluino
are an additional source of OS dileptons.
In cases B1 and B2, where OS dileptons come frequently come
from $\tz_2\to\ell\bar{\ell}\tz_1$ decay, the OS/SS ratio is $\sim6-8$.
Cases B3 and B4 yield intermediate values of this
ratio, in agreement with Fig. 4 and the flavour asymmetry values in Table VII.
Finally, we show in Fig. 5 the fast and slow lepton $p_T$ distributions
for SS events. The hardness of these distributions again reflects the
source of the dileptons. In particular, the slow lepton distribution
in OS dilepton events is typically softer than the corresponding
distribution in SS events. This is reasonable since there is less energy
for multi-leptons produced in the same SUSY particle cascade decay.

\subsection{Three and Four isolated lepton events}

Complicated superparticle cascade decays can lead to events with 3, 4 and
even 5 isolated leptons\cite{BTW} if gluinos are heavy enough.
These cross sections, after the $\eslt > 100$ GeV cut, are shown
in the $3\ell$ and $4\ell$ columns of
Table III. We should warn the reader that in our simulation, we
have just 10-80 trilepton signal events depending on the case
in question and just a handful of $4\ell$ events. These numbers
could, therefore, contain significant statistical fluctuations, and should
be viewed in proper perspective. We see that except in case A2,
$\sim$($10^3-10^4)$ trilepton events should be expected in a data
sample of 10 $fb^{-1}$ at the LHC.
However, since
gluinos are relatively light within our framework, the $4\ell$ cross
sections are rather small.
SM backgrounds to these multi-lepton topologies are small: for our
simulation of 300K $t\bar t$ events, no isolated multi-leptons were found,
leading to an upper bound on BG from this source of $<0.004$ $pb$.
Estimates of backgrounds from $t\bar tt\bar t$ and three and four
vector boson production also yield rates below the $fb$ level\cite{BTW}.

In the $3\ell$ channel, we see a wide range of possible rates for our SUGRA
cases. Case A2, which is dominated by $\tt_1\tt_1$ and $\tg\tg$
production, where $\tg\to t\tt_1$ and $\tt_1\to c\tz_1$, produced no
$3\ell$ or $4\ell$ events, corresponding to an upper limit on this rate of
$\sim 0.02\ pb$. Case A1, which is similar to A2 except now $\tt_1\to b\tw_1$,
can yield substantial $3\ell$ and $4\ell$ rates in gluino pair events
where both tops and stops decay leptonically. Cases B1 and B2, where
$\tz_2\to\ell\bar{\ell}\tz_1$ is common, can also produce large rates
for $3\ell$ and $4\ell$ events. Cases B3 and B4 do not produce large rates
for $3\ell$ or $4\ell$ events due to a small $\tz_2\to\ell\bar{\ell}\tz_1$
branching ratio (see Table II).

\section{Summary and Concluding Remarks}

In this paper, we have studied the implications of the simplest supergravity
grand unified model (based on SU(5) gauge symmetry) with radiative
electroweak symmetry breaking for experiments
at the LHC. This necessarily involves assumptions
about physics at the ultra-high energy scale. Our assumptions,
which include the minimality of the particle content at and
below the GUT scale together with the unification of soft-SUSY
breaking parameters at the unification scale, are carefully elucidated
in Sec. II. In the framework of Minimal Supergravity,
the SUSY parameters at low energy
can all be obtained by renormalization group evolution of just
four GUT scale parameters together with the top quark Yukawa coupling
at the unification scale. We have further restricted the range
of these parameters by incorporating constraints from the non-observation
of proton decay [Eq. (2.3)] and from the requirement that the relic LSP density
is
compatible with the present age of our universe [Eq. (2.4)].
We have taken at face value the implications of the equality of the
$b$-quark and $\tau$-lepton Yukawa couplings at the unification scale
(allowing for 10\% deviations from exact equality due to threshold effects)
which, as has been pointed out by several groups, implies that the
top quark mass is close to its fixed point value.

As discussed in Sec. III,
these requirements lead to a distinct pattern of sparticle masses and mixings.
We find,

\begin{itemize}
\item the gluino is light, $m_{\tg}\leq 400$ GeV;
\item the lighter $t$-squark is substantially lighter than all other squarks,
and
further, $\tb_L$ is lighter than the other squarks;
\item the Higgsino mass parameter $|\mu |$ is large, so that the lighter
charginos and neutralinos are dominantly gauginos, with masses smaller
than about 100 GeV;
\item $\tan\beta$ is close to unity;
\item the lightest scalar in the Higgs sector is never significantly heavier
than 100 GeV.
\end{itemize}

We have seen that these mass and mixing patterns lead to
characteristic features in SUSY events at hadron supercolliders. Of course,
these characteristics can be mimicked by appropriately adjusting the
multitude of parameters
of the MSSM. The point, however, is that an observation of several of
these features could serve as a strong indication of the
simplest unification scenario.

Despite the various theoretical and
experimental constraints that we have imposed, the parameter space
is too large to allow a systematic exploration. Instead, we have
identified six sets of model parameters (that respect all the constraints
in Sec. II) which, we believe, capture the diversity  of signals
that may arise within this framework. The input parameters along with
the sparticle masses are summarized in Table I.
In Cases A1 and A2
the gluino decays via $\tg\to t\tt_1$, while in cases B, gluinos decay
via three body decays. The main difference between A1 and A2 is in
the stop decay patterns: in case A1, $\tt_1\to b\tw_1$ while in case A2,
$\tt_1 \to c\tz_1$.

Our main results for the signals from the supergravity framework are
presented in Sec. V. In order to compare signals from supergravity SU(5)
with those of the usually studied ``minimal models'', we have also
studied the same signals for the four scenarios labelled BTW1-BTW4
defined in Sec. IV, which have a similar value of $m_{\tg}$ as in Table I,
but have squarks and sleptons exactly degenerate. As usual, we
have classified the signals by their leptonic content. Table III
gives an overview of the various signals after all the cuts discussed
in Sec. V. We see that there are observable signals in essentially
all the channels shown including the $1\ell$ and $\ell^+\ell^-$ channels
which are usually thought to be swamped by SM backgrounds.
Unambiguous distinction between all the various
supergravity cases, or for that matter, even between SUGRA and
BTW cases, is not possible, on the basis of cross sections alone.

In order to facilitate this distinction, we have listed the characteristics
of the $\eslt$, $1\ell$ and $\ell^+\ell^-$ event samples in Tables V, VI and
VII,
respectively. A major difference between our supergravity
cases A1, A2, B3 and B4, and the BTW cases,
is that the supergravity events are significantly
richer in $B$ hadrons. Of course, whether distinction of scenarios based
on $B$ identification is actually
possible will crucially depend on the ability of the detectors to
identify displaced vertices. In our calculation, we have assumed
a 40\% efficiency of identifying central $B$ mesons with $p_T > 20$ GeV in the
central region. The reason that supergravity events are $B$-rich is that
one, or more, of
the third generation of squarks is significantly lighter than the
other squarks. In cases A1 and A2 where the gluinos decay via $t\tt_1$,
each gluino pair event contains at least two $b$ quarks from the decays
of the top (in case A1, the decay of each of the stops also gives $b$'s).
In cases
B1-B4, the enrichment of the $B$ sample arises in part
because $\tb_L$ is lighter
than the other squarks. As a result, gluino and neutralino decays
mediated by a virtual $\tb_L$ are enhanced. Furthermore, since $\tz_1$
and $\tz_2$ are mainly gaugino-like, the Z-mediated decays of $\tz_2$
are strongly suppressed. Then, when squarks are very heavy, contributions
from Higgs mediated $\tz_2 \to b\bar{b}\tz_1$ decays, which further
enrich the $b$ sample, become significant. Finally, in cases B2-B4, $B$'s
can also come from $\tg \to \tw_1 t b$ decays, although in case B2,
the primary $b$ quark is likely to be soft. In cases B3 and B4, the
primary $b$ quark and, in all three cases, the $b$ quark from top decay
should have large efficiency to pass our cuts.
BTW type events can be very
rich in $B$'s if $H_{\ell}$ can be produced via cascade decays. As
we have mentioned, this is unimportant for the smaller gluino
masses of interest here (presumably, the kinematic features of
the events will give a rough measure of gluino mass).
We also see from Table V that in cases
A1 and A2, and also in cases B3 and B4, supergravity events can have very
high jet multiplicities---about 10\% of the $\eslt$ sample has $n_j\geq 8$
whereas this number is somewhat smaller for the BTW cases. For cases A1 and
A2, this is due mainly to the production of top quarks in gluino decays,
which leads to a large jet multiplicity. Finally, we
see that the flavour asymmetry $A_f$ defined in Sec. V is generally
larger in the BTW cases than in the SUGRA cases, except when the gluino
is very light. This asymmetry, which is a  measure of the rate for neutralino
production in gluino and squark cascades, is greatly reduced when chargino
cascades dominate as, for instance, in cases A1 and A2.

The features of Tables V-VII also facilitate some distinction between
the various supergravity scenarios. Clearly A1 and A2 are readily
distinguishable
via the high B-multiplicity in the $\eslt$ sample and also from the absence
of any flavour asymmetry in the $\ell^+\ell^-$ sample. The ratio of SS
to OS dilepton production rate in Table III may facilitate further
distinction.
Also, the $B$-multiplicity is
significantly higher in A1 than in A2. The dilepton cross section is
considerably higher in case A1 (where stops can also yield leptons)
than in A2, while exactly the opposite is true for the $\eslt$ signal.
Furthermore, since $\tz_2$ rarely occurs in the A1 and A2 cases, case
A2 yields no $3\ell$ or $4\ell$ cross section, while A1 yields such
events only from top and stop decay.

Distinction amongst the various cases B is somewhat more subtle.
B1 and B2 are very similar, except for the squark mass, as are B3 and B4.
A striking difference between (B1, B2) and (B3,B4) is in the
flavour asymmetry, in the OS/SS ratio in Table III, in
the differential distribution shown in Fig. 4, and especially in the large
rate for isolated $3\ell$ and $4\ell$ events (as has been
explained, these are all measures of $\tz_2$ production via cascade
decays). Some distinction may also be possible via the scalar
$E_T$ distribution since the parent gluino masses are significantly
different. We should mention that the hard scalar $E_T$ cut tends
to wash out this difference; therefore, this difference may best
be studied in the $\eslt$ sample {\it prior} to this cut.
This should be possible since the signal is larger than background
even before this cut as can be seen from Table IV. The multiplicity
of $B$ mesons may provide yet another distinction.
To distinguish
B1 from B2 or B3 from B4 is more difficult.
This is not surprising since essentially the only difference
comes from the squark mass. The clearest distinction
appears to be in the cross section for the $1\ell$ signal and, in
the case of B3 and B4, also from the $\eslt$ cross section. It is
encouraging to see that the various scenarios are distinguishable
even in the messy environment of a hadron collider.

The reader may have noticed that even though the $t$-squark is
light and even though its production constitutes a substantial portion of all
SUSY events (see Table II), we have not extracted any direct signal
for its production. In fact, signals for $\tt_1$-pair
production may even be observable at the Tevatron in several of  the
cases discussed in this paper. Since our primary purpose here was
to study the distinguishability of various scenarios within the
SU(5) supergravity framework from one another and from the so-called
``minimal models'', we used rather hard cuts to obtain
clean event samples.
The direct signals from the lighter sparticles
are essentially removed by these cuts and, in
our analysis, the existence of the lighter stops
and sbottom quarks manifested itself in the decay patterns
of the parent gluinos. It may, of course, well be
possible to devise a strategy that is dedicated to look for the
light scalar top signal at the LHC. Charginos and neutralinos
may also be searched for using specialized strategies\cite{BCPT}.

In summary, we have examined the phenomenology of the minimal SU(5)
supergravity grand unified model with radiative breaking of electroweak
symmetry. We have delineated the range of parameters allowed by
constraints from proton decay and from a cosmologically acceptable
LSP. We have shown that these lead to a characteristic pattern of
sparticle masses and mixings. We have examined signals from the
production of supersymmetric particles at the LHC for representative
points in the parameter space. We find that these signals
are distinguishable from the corresponding signals within the
usually studied models where squarks and sleptons are taken to
be degenerate. We further find that by simultaneously examining various
signals it should be possible to
distinguish between the various scenarios that are possible even
within the restricted framework of supergravity SU(5).
Discovery of supersymmetric signals at the LHC with patterns
corresponding to any of these cases would be extremely dramatic,
since it would not only signal the discovery of supersymmetry,
but would provide support for the ideas underlying supergravity
grand unification.

{\it Note added: As we were completing this manuscript, we received a paper
addressing some of the phenomenological consequences of Yukawa unified
``no-scale'' models\cite{YUMS}.}

%
\acknowledgments

We thank M. Bisset for embedding Higgs radiative corrections into ISAJET,
and R. Arnowitt, P. Nath and F. Paige  for discussions.
This research was supported in part by the U.~S. Department of Energy
under contract numbers DE-FG05-87ER40319, DE-AC02-76ER00881
and DE-AM03-76SF00235.
The work of HB was supported by the TNRLC SSC Fellowship program.
The work of MD was supported in part by
the Wisconsin Research Committee with funds
granted by the Wisconsin Alumni Research Foundation, by the Texas National
Research Laboratory Commission under grant RGFY93--221, as well as by a grant
from the Deutsche Forschungsgemeinschaft under the Heisenberg program.
XT acknowledges support from the US National Science Foundation's
US-Japan Cooperative Research Program.
%
%
%
%

%
\newpage
%
%
%
\begin{table}
\caption[]{Parameters and masses for six SUGRA cases A1, A2 and B1--B4.}
\bigskip
\begin{tabular}{lrrrrrr}
parameter & A1 & A2 & B1 & B2 & B3 & B4 \\
\tableline
$m_0$          & 500 & 500 & 300 & 1000 & 400 & 1000 \\
$m_{1\over 2}$ & 120 & 130 & 60 & 70 & 130 & 130 \\
$A_0/m_0$      & 0.6 & 3.75 & 0.1 & 0.0 & 0.0 & 0.0 \\
$\tan\beta$    & 1.49 & 2.2 & 1.94 & 1.32 & 2.22 & 2.11 \\
$\mu$          & 697.1 & 580 & -313.3 & -1571.7 & 430 & 964.7 \\
$m_t$          & 160 & 175 & 170 & 155 & 175 & 175 \\
$m_{\tg}$      & 346 & 371 & 185 & 231 & 364 & 400 \\
$m_{\tq}$      & 568 & 578 & 328 & 1011 & 496 & 1039 \\
$m_{\tt_1}$    & 131 & 83.6 & 198 & 121 & 225 & 288 \\
$m_{\tt_2}$    & 521 & 500 & 304 & 781 & 470 & 799 \\
$m_{\tb_L}$    & 437 & 426 & 250 & 732 & 396 & 765 \\
$m_{\tl}$      & 508 & 501 & 305 & 1001 & 412 & 1005 \\
$m_{\tz_1}$    & 43.6 & 48.0 & 27.1 & 28.8 & 46.1 & 50.7 \\
$m_{\tz_2}$    & 85.1 & 93.5 & 63.5 & 59.8 & 89.3 & 99.9 \\
$m_{\tz_3}$    & 698 & 582 & 316 & 1572 & 432 & 966 \\
$m_{\tz_4}$    & 710 & 595 & 333 & 1577 & 450 & 973 \\
$m_{\tw_1}$    & 84.5 & 92.9 & 62.2 & 59.7 & 87.9 & 99.6 \\
$m_{\tw_2}$    & 707  & 593 & 331   & 1576 & 448 & 972 \\
$m_{H_\ell}$   & 85.2 & 99.7 & 61.6 & 86.0 & 91.3 & 101 \\
$m_{H_h}$      & 1038 & 841 & 492 & 2339 & 652 & 1543 \\
$m_{H_p}$      & 1038 & 842 & 487 & 2342 & 649 & 1542 \\
$m_{H^\pm}$    & 1040 & 845 & 492 & 2343 & 653 & 1544 \\
\end{tabular}
\end{table}
\begin{table}
\caption[]{(a) Fractions of SUSY particle pairs produced in $pp$ collisions
at the LHC; and (b) branching fractions of selected decay modes,
for six SUGRA cases A1, A2 and B1--B4, where $\tq$ stands for all squarks
except stops, and $\tx\tx$ stands for all possible
chargino and neutralino pairs: $\tw_i\tw_j$, $\tz_i\tz_j$, and $\tw_i\tz_j$.}
\bigskip
\begin{tabular}{lllllll}
SUSY particles$\backslash$ Case & A1 & A2 & B1 & B2 & B3 & B4 \\
\tableline
(a) SUSY Particle Pairs Produced \\
$\tg\tg$         & 0.30  & 0.093 & 0.72  & 0.74  & 0.44  & 0.73  \\
$\tt\tt$         & 0.51  & 0.84  & 0.011 & 0.21  & 0.10  & 0.083 \\
$\tg\tq$         & 0.13  & 0.050 & 0.23  & 0.013 & 0.33  & 0.081 \\
$\tq\tq$         & 0.018 & 0.007 & 0.029 & 3$\times 10^{-4}$ & 0.067 & 0.005 \\
$\tw_1^\pm\tz_2$ & 0.018 & 0.006 & 0.004 & 0.019 & 0.027 & 0.066 \\
$\tx\tx$         & 0.026 & 0.009 & 0.007 & 0.027 & 0.042 & 0.088 \\
(b) Important Decay Modes \\
$\tg\to \tt_1 \bar{t},\bar{\tt_1}t$   & 1.0 & 1.0 & - & - & - & - \\
$\tg\to \tw_1^-t\bar{b},\tw_1^+b\bar{t}$ & 1.6$\times 10^{-4}$ & 1.4$\times
10^{-4}$
                         & -     & 0.091 & 0.12 & 0.25 \\
$\tg\to \tw_1^-u\bar{d},\tw_1^+d\bar{u}$ & 4.6$\times 10^{-4}$ & 2.9$\times
10^{-4}$
                         & 0.21  & 0.10 & 0.19 & 0.16 \\
$\tg\to \tz_1 d\bar{d}$   & 3.1$\times 10^{-5}$ & 1.9$\times 10^{-5}$
                         & 0.012 & 0.005 & 0.014 & 0.01 \\
$\tg\to \tz_1b\bar{b}$   & 6.8$\times 10^{-5}$ & 5.4$\times 10^{-5}$
                         & 0.012 & 0.006 & 0.035 & 0.018 \\
$\tg\to \tz_2b\bar{b}$   & 4.3$\times 10^{-4}$ & 3.9$\times 10^{-4}$
                         & 0.21  & 0.10  & 0.18  & 0.15 \\
$\tt_1\to \tw_1^+ b$     & 1.0 & -   & 0.95 & 1.0 & 0.93 & 0.29 \\
$\tt_1\to \tz_1 t$       & -   & -   & 0.05 & -   & 0.07 & 0.64 \\
$\tt_1\to \tz_2 t$       & -   & -   & -    & -   & -    & 0.07 \\
$\tt_1\to \tz_1 c$       & -   & 1.0 & -    & -   & -    & -    \\
$\tz_2\to\tz_1 d\bar{d}$ & 0.11  & 0.18   & 0.024
                         & 0.028  & 0.21   & 0.17 \\
$\tz_2\to\tz_1 b\bar{b}$ & 0.37  & 0.39   & 0.057
                         & 0.22  & 0.38   & 0.42 \\
$\tz_2\to\tz_1 e^+e^-$   & 0.047 & 0.018  & 0.14
                         & 0.12  & 0.007  & 0.012 \\
$\tw_1^+\to\tz_1 e^+\nu_e $ & 0.11 & 6.9$\times 10^{-5}$
                            & 0.11 & 0.11 & 0.11 & 0.11
\end{tabular}
\end{table}

\begin{table}
\caption[]{Cross sections in $pb$ for various event topologies after cuts
described in the text, for $pp$ collisions at $\sqrt{s}=14$~TeV.
The various SUGRA cases are listed in the first column. The OS/SS ratio
is computed with the OS dilepton sample {\it before} the scalar $E_T$ cut.}

\bigskip

\begin{tabular}{cccccccc}
case & $\eslt$ & $1\ \ell$ & $OS$ & $SS$ & OS/SS & $3\ \ell$ & $4\ \ell$ \\
\tableline
A1 & 24.6 & 36.2 & 5.4 & 3.7 & 2.0 & 1.2 & 0.017 \\
A2 & 48.0 & 31.4 & 1.5 & 2.1 & 1.2 & $<0.02$ & $<0.02$ \\
B1 & 79.1 & 76.8 & 11.9 & 3.4 & 6.9 & 1.7 & 0.17 \\
B2 & 67.7 & 37.5 & 9.0 & 2.4 & 7.7 & 0.8 & 0.1 \\
B3 & 51.8 & 21.2 & 1.4 & 0.6 & 3.3 & 0.09 & $<0.01$ \\
B4 & 20.1 & 10.1 & 1.1 & 0.4 & 3.5 & 0.1 & $<0.004$ \\
BTW1 & 105 & 39.8 & 2.8 & 1.4 & 3.1 & 0.21 & 0.03 \\
BTW2 & 57.3 & 22.5 & 2.2 & 0.85 & 3.6 & 0.14 & $<0.02$ \\
BTW3 & 96 & 58 & 10.9 & 2.9 & 6.7 & 1.5 & 0.06 \\
BTW4 & 52.3 & 23.9 & 2.3 & 0.9 & 3.7 & 0.2 & $<0.02$ \\
$t\bar t(160)$ & 2.9 & 8.0 & 1.1 & 0.01 & 640 & $<0.004$ & $<0.004$ \\
$W+jet$ & 0.6 & 3.8 & 0.29 & -- & & -- & -- \\
$WW$ & -- & -- & 0.001 & -- & & -- & -- \\
$Z+jet$ & 0.6 & 0.2 & 0.02 & -- & & -- & -- \\
$b\bar b,c\bar c$ & 0.2 & 0.02 & -- & -- & & -- & -- \\
$total\ BG$ & 4.3 & 12.02 & 1.411 & 0.01 & & $<0.004$ & $<0.004$ \\
\end{tabular}
\end{table}
\begin{table}
\caption[]{Cross sections in pb for various event topologies (before) after
the cuts on scalar $E_T$ and $M_T$ described in Sec. V of the text,
for $pp$ collisions at $\sqrt{s}=14$~TeV.
The various SUGRA cases are listed in the first column.}

\bigskip

\begin{tabular}{cccc}
case & $\eslt$ & $1\ \ell$ & $OS$ \\
\tableline
A1 & (26.3) 24.6 & (62.6) 36.2 & (7.3) 5.4 \\
A2 & (53.9) 48.0 & (50.1) 31.4 & (2.5) 1.5 \\
B1 & (103) 79.1 & (209) 76.8 & (23.5) 11.9 \\
B2 & (93.0) 67.7 & (122) 37.5 & (18.6) 9.0 \\
B3 & (56.8) 51.8 & (35.0) 21.2 & (2.0) 1.4 \\
B4 & (21.7) 20.1 & (16.3) 10.1 & (1.4) 1.1 \\
BTW1 & (127) 105 & (85.5) 39.8 & (4.5) 2.8 \\
BTW2 & (65.4) 57.3 & (42.0) 22.6 & (3.1) 2.2 \\
BTW3 & (119) 96.0 & (127) 58.1 & (19.3) 10.9 \\
BTW4 & (58.9) 52.3 & (43.3) 23.9 & (3.2) 2.3 \\
$t\bar t(160)$ & (4.6) 2.9 & (52.4) 8.0 & (6.4) 1.1 \\
$W+jet$ & (1.7) 0.6 & (108) 3.8 & (3.8) 0.29 \\
$Z+jet$ & (1.2) 0.6 & (1.4) 0.2 & (0.10) 0.02 \\
$b\bar b,c\bar c$ & (0.24) 0.20 & (0.13) 0.02 & -- (--) \\
$total\ BG$ & (7.7) 4.3 & (161.9) 12.02 & 10.3 (1.41) \\
\end{tabular}
\end{table}
\begin{table}
\caption[]{Cross sections and event fractions for missing
energy plus jets events for various SUSY cases
at the LHC.}

\bigskip

\begin{tabular}{ccccccccc}
case & $\sigma (pb)$ & $n_j=4-5$ & $n_j=6-7$ & $n_j\ge 8$ &
$n_b\ge 1$ & $n_b\ge 2$ & $<\Sigma E_T>$ & $<\eslt >$ \\
\tableline
A1 & 24.6 & 0.54 & 0.35 & 0.10 & 0.65 & 0.28 & 1160 & 212 \\
A2 & 48.0 & 0.55 & 0.35 & 0.10 & 0.47 & 0.09 & 1112 & 221 \\
B1 & 79.1 & 0.73 & 0.25 & 0.02 & 0.21 & 0.04 & 964 & 195 \\
B2 & 67.7 & 0.77 & 0.20 & 0.03 & 0.23 & 0.05 & 999 & 211 \\
B3 & 51.8 & 0.57 & 0.35 & 0.08 & 0.36 & 0.12 & 1118 & 215 \\
B4 & 20.1 & 0.54 & 0.36 & 0.10 & 0.44 & 0.15 & 1204 & 217 \\
BTW1 & 105 & 0.69 & 0.26 & 0.04 & 0.17 & 0.04 & 1006 & 214 \\
BTW2 & 57.3 & 0.61 & 0.33 & 0.05 & 0.18 & 0.04 & 1091 & 211 \\
BTW3 & 96 & 0.69 & 0.27 & 0.04 & 0.15 & 0.03 & 1011 & 217 \\
BTW4 & 52.3 & 0.60 & 0.33 & 0.06 & 0.18 & 0.04 & 1109 & 208 \\
$t\bar t(160)$ & 2.9 & 0.81 & 0.17 & 0.01 & 0.56 & 0.12 & 895 & 201 \\
$Z+jets$ & 0.63 & 0.89 & 0.11 & 0.0 & 0.11 & 0.02 & 905 & 281 \\
\end{tabular}
\end{table}

\begin{table}
\caption[]{Cross sections and event fractions for events
containing a single lepton plus jets for various SUSY cases
at the LHC.}

\bigskip

\begin{tabular}{ccccccccc}
case & $\sigma (pb)$ & $n_j=0-3$ & $n_j=4-5$ & $n_j\ge 6$ &
$n_b\ge 1$ & $n_b\ge 2$ & $<\Sigma E_T>$ & $<\eslt >$ \\
\tableline
A1 & 36.2 & 0.23 & 0.49 & 0.28 & 0.67 & 0.29 & 1056 & 167 \\
A2 & 31.4 & 0.26 & 0.54 & 0.21 & 0.45 & 0.10 & 1020 & 191 \\
B1 & 76.8 & 0.35 & 0.50 & 0.15 & 0.27 & 0.06 & 913 & 148 \\
B2 & 37.5 & 0.41 & 0.50 & 0.09 & 0.23 & 0.04 & 910 & 165 \\
B3 & 21.2 & 0.24 & 0.53 & 0.24 & 0.36 & 0.1 & 1052 & 178 \\
B4 & 10.1 & 0.22 & 0.52 & 0.26 & 0.46 & 0.16 & 1108 & 185 \\
BTW1 & 39.8 & 0.32 & 0.52 & 0.15 & 0.15 & 0.03 & 970 & 176 \\
BTW2 & 22.6 & 0.26 & 0.54 & 0.19 & 0.15 & 0.03 & 1030 & 172 \\
BTW3 & 58 & 0.33 & 0.52 & 0.15 & 0.15 & 0.03 & 969 & 174 \\
BTW4 & 23.9 & 0.24 & 0.55 & 0.20 & 0.15 & 0.04 & 1024 & 168 \\
$t\bar t(160)$ & 8 & 0.49 & 0.46 & 0.04 & 0.52 & 0.12 & 857 & 153 \\
$W+jets$ & 3.8 & 0.92 & 0.08 & 0.0 & 0.08 & 0.0 & 927 & 208 \\
\end{tabular}
\end{table}

\begin{table}
\caption[]{Cross sections and event fractions for events
containing opposite sign dileptons plus jets for various SUSY cases
at the LHC.}

\bigskip

\begin{tabular}{cccccccccc}
case & $\sigma (pb)$ & $n_j=0-3$ & $n_j=4-5$ & $n_j\ge 6$ &
$n_b\ge 1$ & $n_b\ge 2$ & $<\Sigma E_T>$ & $<\eslt >$ &
$A_f$ \\
\tableline
A1 & 5.4 & 0.33 & 0.47 & 0.20 & 0.69 & 0.32 & 1017 & 165 & $-0.009\pm 0.06$ \\
A2 & 1.5 & 0.39 & 0.47 & 0.14 & 0.47 & 0.11 & 950 & 161 & $0.05\pm 0.16$ \\
B1 & 11.9 & 0.54 & 0.41 & 0.06 & 0.35 & 0.14 & 895 & 135 & $0.77\pm 0.07$ \\
B2 & 9.0 & 0.42 & 0.51 & 0.07 & 0.26 & 0.07 & 920 & 163 & $0.79\pm 0.05$ \\
B3 & 1.4 & 0.29 & 0.50 & 0.21 & 0.53 & 0.15 & 1035 & 179 & $0.32\pm 0.08$ \\
B4 & 1.1 & 0.30 & 0.54 & 0.16 & 0.58 & 0.25 & 1050 & 183 & $0.26\pm 0.05$ \\
BTW1 & 2.8 & 0.41 & 0.50 & 0.08 & 0.20 & 0.03 & 942 & 166 & $0.7\pm 0.07$ \\
BTW2 & 2.2 & 0.29 & 0.52 & 0.19 & 0.23 & 0.03 & 1042 & 162 & $0.6\pm 0.07$ \\
BTW3 & 10.9 & 0.39 & 0.51 & 0.10 & 0.17 & 0.03 & 984 & 165 & $0.78\pm 0.03$ \\
BTW4 & 2.3 & 0.36 & 0.50 & 0.14 & 0.21 & 0.03 & 990 & 156 & $0.77\pm 0.05$ \\
$t\bar t(160)$ & 1.1 & 0.60 & 0.38 & 0.02 & 0.50 & 0.12 & 827 & 144 &
$0.01\pm 0.06$ \\
\end{tabular}
\end{table}


%
\begin{figure}
\caption[]{Transverse momentum distribution of fastest jet in
multi-jet plus $\eslt$ events for four SUGRA cases and the sum of
backgrounds (histogram), at $\sqrt{s}=14$ TeV.}
\end{figure}
%
\begin{figure}
\caption[]{Distributions for {\it a}) scalar $E_T$, {\it b}) lepton plus
missing energy transverse mass, and {\it c}) $p_T(lepton)$ in single
isolated lepton events for four SUGRA cases A1 (dashes), A2 (dot-dash),
B1 (double-dot-dash) and B4 (triple-dot-dash), along with SM background
(solid histogram), at $\sqrt{s}=14$ TeV. Distributions {\it a}) and {\it b})
are made before scalar $E_T$ and $m_T$ cuts, while {\it c}) is after cuts.}
\end{figure}
%
\begin{figure}
\caption[]{Distributions for {\it a}) scalar $E_T$, {\it b}) fast lepton
transverse momentum, and {\it c}) slow lepton transverse momentum in
opposite sign isolated dilepton events for four SUGRA
cases A1 (dashes), A2 (dot-dash),
B2 (double-dot-dash) and B4 (triple-dot-dash), along with SM background
(solid histogram), at $\sqrt{s}=14$ TeV. Distribution {\it a})
is made before the scalar $E_T$ cut, while {\it b}) and {\it c}) are
after the scalar $E_T$ cut.}
\end{figure}
\begin{figure}
\caption[]{Distributions for dilepton opening angle in the transverse
plane for OS dileptons and SS dileptons for SUGRA cases A1, B2 and B4.}
\end{figure}
%
\begin{figure}
\caption[]{Distributions for {\it a}) fast lepton
transverse momentum, and {\it b}) slow lepton transverse momentum in
same sign isolated dilepton events for four SUGRA
cases A1 (dashes), A2 (dot-dash),
B2 (double-dot-dash) and B4 (triple-dot-dash), along with SM background
(solid histogram), at $\sqrt{s}=14$ TeV.}
\end{figure}
%
%


\begin{references}
\bibitem{FN1} We do not consider the large $\tan\beta$ solution corresponding
to the fixed point of the bottom Yukawa interaction because radiative
breaking constraints require that $\tan\beta \leq m_t/ m_b$.
\bibitem{FN2} R.~Arnowitt and P.~Nath, Phys. Rev. Lett. {\bf69},
725 (1992), which takes into account the constraints from radiative symmetry
breaking, have concluded that the non-observation of proton decay implies
the bound $m_{\rm 0}\geq 500$ GeV; while we have not incorporated
their more restrictive analysis into our considerations, we mention
that this bound is respected by four of the six cases that we study in
the following.
\bibitem{FN3} The arguments for not considering the mass ratios for the
$T_3 = -1/2$ fermions for the first two generations are the same as for
standard SU(5). We also remind the reader that we have allowed for some
variation of $m_b/m_{\tau}$ to allow for threshold effects and effects
of suppressed non-renormalizable operators.
\bibitem{FN4} The incorporation of the pole gluino mass can substantially
change the phenomenology especially when the gluino is very light.
\end{references}
\end{document}